\documentclass{ws-ijmpd}

\usepackage{graphicx}
\usepackage{dcolumn}
\usepackage{array,multirow}
\usepackage{bm}
\usepackage{amsmath}
\usepackage{epstopdf}
\usepackage{amsfonts}
\usepackage{amssymb}
\usepackage{epsfig}
\usepackage{tabularx}
\usepackage{color}
\usepackage[colorlinks=true,citecolor=blue,urlcolor=magenta,breaklinks]{hyperref}
\usepackage{hyperref}
\usepackage{cite}

\newcommand{\be}{\begin{equation}}
\newcommand{\en}{\end{equation}}
\newcommand{\bea}{\begin{eqnarray}}
\newcommand{\ena}{\end{eqnarray}}

\begin{document}


%
\catchline{}{}{}{}{}
%

\title{Millisecond pulsars modelled as strange quark stars
	   admixed with condensed dark matter}

\author{Grigoris Panotopoulos}

\address{Centro de Astrof\'{\i}sica e Gravita{\c c}{\~a}o, Instituto Superior T\'ecnico-IST,
Universidade de Lisboa-UL, Av. Rovisco Pais, 1049-001 Lisboa, Portugal
\\
\href{mailto:grigorios.panotopoulos@tecnico.ulisboa.pt}{\nolinkurl{grigorios.panotopoulos@tecnico.ulisboa.pt}} 
}

\author{Il{\'i}dio Lopes}

\address{Centro de Astrof\'{\i}sica e Gravita{\c c}{\~a}o, Instituto Superior T\'ecnico-IST,
Universidade de Lisboa-UL, Av. Rovisco Pais, 1049-001 Lisboa, Portugal
\\
\href{mailto:ilidio.lopes@tecnico.ulisboa.pt}{\nolinkurl{ilidio.lopes@tecnico.ulisboa.pt}} }

\maketitle

\begin{history}
\received{Day Month Year}
\revised{Day Month Year}
\end{history}

\begin{abstract}
We study for the first time how a new class of stars could impact an ensemble of pulsars with known masses and spin-periods. These new compact objects are strange stars admixed with condensed dark matter. In this exploratory theoretical work, our goal is to determine how the basic parameters of pulsars are modified for such a new class of compact objects. In particular we consider three different scenarios that correspond to a dark matter mass fraction of $~5\%$, $~11\%$ and $~25\%$. Within each scenario with fixed parameters we predict theoretically other properties of the pulsars, such as the radius, the compactness, the moment of inertia, as well as the angular momentum. Our numerical results are summarized in tables and also shown graphically for better visualization, where a comparison between the different scenarios can be made. 
\end{abstract}


\ccode{}

\section{Introduction}

After the pioneer work of F. Zwicky in 1933 \cite{zwicky}, and much later of V. Rubin in 1970 \cite{rubin}, we realized that most of the non-relativistic matter in the Universe is dominated by dark matter (DM). Recent, well established data from Astrophysics and Cosmology indicate that we live in a spatially flat Universe dominated by dark energy, and also confirm and support the existence of DM \cite{turner}. Dark energy and dark matter comprise two of the biggest challenges of modern Cosmology since their origin and nature still remains a mystery. In this article we shall be discussing dark matter. For a review on this topic see e.g. \cite{DM1,DM2}, and for recent ones on direct, indirect and collider DM searches see e.g. \cite{DM3,DM4,DM5}.

If DM consists of bosons, then DM inside a star may be modelled as a Bose-Einstein condensate, leading either to a dark matter star \cite{darkstars} or to compact stars admixed with condensed dark matter. Even if ordinary matter and DM only interact indirectly via gravity, DM can have significant gravitational effects on compact objects, which have been investigated in
\cite{boson1,boson2,boson3,boson4} for bosonic condensed DM, and in \cite{fermion1,fermion2,fermion3,fermion4,fermion5} for fermionic DM.
Alternative scenarios for which DM is made of bosons can be found in~\cite{britoetal1,britoetal2}. 

Compact objects \cite{textbook}, such as white dwarfs and neutron stars, are the final fate of stars, and comprise excellent cosmic laboratories to study, test and constrain new physics and/or gravitational theories under extreme conditions that cannot be reached in Earth-based experiments. It is well-known that the properties of compact objects, such as mass and radius, depend crucially on the equation of state of ultra-dense matter which unfortunately is poorly known. Soon after the discovery of the neutron by James Chadwick, Baade and Zwicky predicted that neutron stars should exist \cite{baade}. One year after the discovery of pulsars in 1967, their identification as neutron stars was established after the discoveries of pulsars in the Crab and Vela supernova remnants \cite{chamel}. Recently a new class of compact objects has
been postulated to exist due to some observed super-luminous supernovae \cite{SL1,SL2}, which occur in about one out of
every 1000 supernovae explosions, and are more than 100 times brighter than normal supernovae. One plausible explanation is that neutrons are
further compressed so that a new object made of de-confined quarks is formed. This new compact object is called a "strange quark star" \cite{SS1,SS2,SS3}, and since it is a much more stable configuration compared to a neutron star, it could explain the origin of the huge amount of energy released in super-luminous supernovae. In this short article we make a first study about the possibility of such a class of objects being among known pulsars.

The Milky Way is known to be populated by a large population of compact objects, such as  white dwarfs, neutron stars and possibly even stellar black holes~\cite{ilidio17a,ilidio17b}. Many of these compact objects are know to be neutron stars. It is estimated that the Milky Way population of neutron stars is  around one billion, of which probably 200 000 are pulsars. Until now astronomers have discovered  slightly less than 2 000 pulsars.
Therefore, there is the distinct possibility that among the many compact stars that exist in our Galaxy, some of them could be unknown objects made of exotic matter like strange quark stars. Moreover, since the Milky Way, as any other spiral galaxy, is made of more than 90\% of DM \cite{milkyway}, it is possible that some of these stars could be made of large amounts of DM. In particular,
we are interested in looking for the impact of DM in strange quark stars, some of which could be part of the known population of pulsars. As such, we will regard a small set of pulsars as possible candidates for strange quark stars with dark matter (see table ~\ref{table:SetPulsars}). In this preliminary work we estimate how the usual observation parameters of pulsars will be different for such a class of hypothetical objects. We notice that such a set of well-known pulsars was chosen in our study only for illustrative purposes.
       
This work is substantiated by the fact that there is some evidence that strange stars may exist. Indeed the recent discovery of very  compact objects (with very high densities)  like  the  millisecond pulsars SAX J 1808.4-3658 and RXJ185635-3754, the X-ray burster 4U 1820-30, the X-ray pulsar Her X-1, and  X-ray source PSR 0943+10, are among the best candidates \cite{lattimer}. Moreover, the recently launched NASA mission NICER, designed primarily to observe thermal X-rays emitted by several millisecond pulsars, could help answer to this question \cite{nicer}.

As it was pointed out in \cite{basic}, the recent fast growth of millisecond pulsars with precisely measured mass provides us with an excellent opportunity to probe the physics of compact stars, since the stellar parameter values can be accurately computed
for known mass and spin rate, on the one hand, and a given equation of state for the ultra-dense matter inside the star, on the other hand. The authors of \cite{basic} provided the first detailed catalogue of numerically computed parameter values for 16 observed pulsars,  assuming 8 different equations of state corresponding to nucleonic, hyperonic, hybrid and strange matter. It is the aim of our  article to study the effects of bosonic condensed DM on the properties of observed pulsars with known spin period and mass.

Our work is organized as follows: 
after this introduction, we present the equations of state in section two, while the equations for hydrodynamical equilibrium are presented in the third section. Our numerical results are discussed
in section four, and finally we conclude in the last section.
We use metric signature (-,+,+,+), and we work in natural units in which $c=1=\hbar$.
In these units all dimensionful quantities are measured in GeV, and we make use of the conversion rules $1 m = 5.068 \times 10^{15} GeV^{-1}$ and $1 kg = 5.610 \times 10^{26} GeV$ \cite{guth}.

\section{Equations of state}

\subsection{Ordinary quark matter}

For strange matter we shall consider the simplest equation of state corresponding to a relativistic gas of de-confined quarks,
known also as the MIT bag model \cite{MIT1,MIT2,MIT3}
\be
P_s = \frac{1}{3} (\epsilon_s - 4B)
\en
and the bag constant has been taken to be $B^{1/4}=148 MeV$ \cite{Bvalue}.
Although refinements of the bag model exist in
the literature \cite{refine1,refine2,refine3} (for the present state-of-the-art 
see the recent paper \cite{art}), the above analytical expression "radiation plus constant" has been employed in recent works, both in GR \cite{prototype}, where it was shown that the observed value of the cosmological constant $\Lambda \sim (10^{-33} eV)^2$ is too small to have an effect, and in R-squared gravity \cite{kokkotas1,kokkotas2}. Therefore, despite its simplicity we will consider the MIT bag model in the present work, since it suffices for our purposes, and we will take the cosmological constant to be zero.

\begin{table}[ht!]
\tbl{List of observed pulsars discussed in this work (from~\cite{basic}).}
{
\begin{tabular}{l | l l l}
		No & Pulsar name & Spin-period (ms) & Mass ($M_{\odot}$) \\
		\hline
		\hline
		1 & J1918-0642 & 7.60 & 1.18 \\
		2 & J1738+0333 & 5.85 & 1.47 \\
		3 & J1012+5307 & 5.26 & 1.83 \\
		4 & J0751+1807 & 3.48 & 1.64 \\
		5 & B1855+09 & 5.36 & 1.30
\end{tabular}
	\label{table:SetPulsars}
}
\end{table}

\subsection{Condensed dark matter}

The DM particles are usually assumed to be collisionless. However in \cite{self-interacting} the authors introduced the idea that
DM may have self interactions in order to alleviate some apparent conflicts between the collisionless cold dark matter paradigm and astrophysical
observations. It was found in \cite{self-interacting} that the appropriate range for the strength of self-interaction has to be
$ 0.45\, cm^2g^{-1} < \sigma_\chi/m_\chi < 450\, cm^2g^{-1} $
where $m_\chi$ is the mass of the DM particles, and $\sigma_\chi$ is the self interaction cross section of dark matter. Current limits on the strength
of the dark matter self interaction read \cite{bullet1,bullet2,review}
$1.75 \times 10^{-4} \, cm^2g^{-1} < \sigma_\chi/m_\chi < (1-2) \, cm^2g^{-1} $.

We model DM inside a star as a strongly-coupled dilute cold boson gas. Under these conditions only binary collisions at low energy are relevant, and thus they are characterized by the s-wave scattering length $l$ irrespectively of the details of the two-body potential \cite{darkstars}. Therefore, the ground state properties of DM are described by the mean-field Gross-Pitaevskii equation \cite{BEC1,BEC2}, also known as non-linear Schr{\"o}dinger equation, with a short range repulsive delta-potential of the form
\begin{equation}
V(\vec{r}_1-\vec{r}_2) = \left( \frac{4 \pi l}{m_\chi} \right) \delta^{(3)}(\vec{r}_1-\vec{r}_2)
\end{equation}
which implies a dark matter self interaction cross section given by $\sigma_\chi=4 \pi l^2$ \cite{darkstars}. Almost all DM particles are in the condensate, the effective pressure of which is computed to be \cite{darkstars}
\begin{equation}
P_\chi = \left( \frac{2 \pi l}{m_\chi^3} \right) \epsilon_\chi^2=K \epsilon_\chi^2 
\end{equation}
Assuming a scattering length $l=(\textrm{a few})~fm$ and a mass $m_\chi = (0.1-5) GeV$, the bounds on $\sigma_\chi/m_\chi$ mentioned before are satisfied, and in the following we shall assume for the constant $K$ the values:
$ K = 1.01 B^{-1} $ for the first admixed scenario (hereafter $A_1$ model),
$ K = 0.46 B^{-1} $ for the second admixed scenario (hereafter $A_2$ model), $ K = 0.5 B^{-1} $, for the third admixed scenario (hereafter $A_3$ model), and finally $ K = 0.02 B^{-1} $ for the pure DM star (hereafter $D$ model). The strange star without DM is called the $S$ model.
Table~\ref{table:SetDMstars} presents the ensemble of stellar scenarios discussed  in this work.

\begin{table}[ht!]
\tbl{List of five distinct scenarios considered in this work.}
{
\begin{tabular}{l | | l l l l}
		No & Model &DM & $f$  & Description \\
		\hline
		\hline
		1  & $S$     & 0\% &-  & Strange star \\
		2 & $A_1$  & 5\% &0.09 & Admixed DM star\\
		3 & $A_2$  & 11\%&0.2 & Admixed  DM star \\
		4 & $A_3$  & 25\%&0.9 & Admixed  DM star \\
		5 & $D$   & 100\% &-  & Condensed DM star
\end{tabular}
	\label{table:SetDMstars}
}
\end{table}

\section{Hydrostatic equilibrium}

Starting from Einstein's field equations, and assuming static spherically symmetric solutions
\be
ds^2 = -e^{\nu(r)} dt^2 + e^{\lambda(r)} dr^2 + r^2 d \Omega^2,
\en
where $e^{\lambda(r)}=(1-2m(r)/r)^{-1}$, one obtains the
Tolman-Oppenheimer-Volkoff (TOV) equations \cite{TOV} for the interior solution of a relativistic star
\bea
m'(r) & = & 4 \pi r^2 \epsilon(r), 
\ena
\bea
P'(r) & = & - (P(r)+\epsilon(r)) \: \frac{m(r)+4 \pi P(r) r^3}{r^2 (1-\frac{2 m(r)}{r})}
\label{eq:P}
\ena
and
\bea
\nu'(r) & = & -2 \; \frac{m(r)+4 \pi P(r) r^3}{r^2 (1-\frac{2 m(r)}{r})} ,
\ena
where the prime denotes differentiation with respect to r. The first two equations are to be integrated with the initial conditions
$m(r=0)=0$ and $P(r=0)=P_c$, where $P_c$ is
the central pressure. The radius of the star is determined requiring that the energy density vanishes at the surface,
$P(R) = 0$, and the mass of the star is then given by $M=m(R)$. Finally, the other metric function can be computed using the third equation
together with the boundary condition $\nu(R)=ln(1-2M/R)$.

Now let us assume that the star consists of two fluids, namely strange matter (de-confined quarks) and dark matter, with only gravitational interaction between
them, and equations of state $P_s(\epsilon_s)$, $P_\chi(\epsilon_\chi)$, respectively. The total pressure and the total energy density of the system are given by $P=P_s+P_\chi$ and $\epsilon=\epsilon_s+\epsilon_\chi$, respectively.
Since the energy momentum tensor of each fluid is separately conserved, the TOV equations in the two-fluid formalism for the interior solution
of a relativistic star with a vanishing cosmological constant  \cite{2fluid1,2fluid2}. Accordingly, equation~\ref{eq:P} can be explicit for
each component of matter: 
\bea
P_s'(r) & = & - (P_s(r)+\epsilon_s(r)) \: \frac{m(r)+4 \pi P(r) r^3}{r^2 (1-\frac{2 m(r)}{r})} 
\ena
and 
\bea
P_\chi'(r) & = & - (P_\chi(r)+\epsilon_\chi(r)) \: \frac{m(r)+4 \pi P(r) r^3}{r^2 (1-\frac{2 m(r)}{r})}.
\ena
and also the equations for the mass function of the two species
\begin{eqnarray}
m_s'(r) & = & 4 \pi r^2 \epsilon_s(r) \\
m_{\chi}'(r) & = & 4 \pi r^2 \epsilon_\chi(r) \\
m(r) & = & m_s(r) + m_{\chi}(r)
\end{eqnarray}
In this case in order to integrate the TOV equations we need to specify the central values both for normal matter and for
dark matter $P_s(0)$ and $P_\chi(0)$, respectively. So we define the dark matter fraction as follows
\be
f = \frac{P_\chi(0)}{P_s(0)+P_\chi(0)}
\en
and we shall consider three different numerical values, namely $f=0.09, 0.2, 0.9$. We have chosen these values in agreement with the current dark matter constraints obtained from compact stars, main sequence stars and the Sun \cite{ref1,ref2,ref3,ref4,ref5,ref6,ref7}.
Actually, as shown by several authors, even smaller amounts of DM (as a percentage of the total mass of the star) can have a quite visible impact on the structure of stars \cite{silk,ilidio2,ilidio5}. Nevertheless, to make a comprehensive study of the impact of this type of DM in the structure of strange quark stars, we analyse stars with different amounts of DM, as well as stars made of 100\% DM or 0 \% DM (cf. Table~\ref{table:SetDMstars}).

\begin{table}[ht!]
\tbl{Models for known pulsars.}
{
	\resizebox{8cm}{!}{  	
		\begin{tabular}{l |l l l l l l l}
			Pulsar Data &  & Model  & $R$  & $\beta$ & $I$ & $J$ \\
			{\scriptsize {\bf M}\,(M$_{\odot}$)\;{\bf f}\,(Hz)} &&& {\scriptsize $(Km)\;$} &   &
			{\scriptsize $(10^{45} g cm^2)$} &  {\scriptsize $(10^{48} g cm^2 s^{-1})$} \\
			\hline
			\hline
			\multirow{1}{*}{{J1918-0642}}
			&&$S$   & 10.19 & 0.17 & 1.08 & 0.89 \\
			\multirow{1}{*}{\bf 1.18   131.6}
			&&$A_1$ & 10.12 & 0.17 & 1.07 & 0.89 \\
			&&$A_2$ & 10.05 & 0.18 & 1.06 & 0.88 \\
			&&$A_3$ & 10.58 & 0.17 & 1.18 & 0.97 \\
			&&$D$   & 16.96 & 0.10  & 1.99  & 1.65\\
			\\
			\multirow{1}{*}{{B1855+09}}
			&& $S$   & 10.43 & 0.19 & 1.26 & 1.48 \\
			\multirow{1}{*}{\bf 1.30  186.6}
			&&$A_1$ & 10.37 & 0.19 & 1.24 & 1.46 \\
			&&$A_2$ & 10.29 & 0.19 & 1.23 & 1.44 \\
			&&$A_3$ & 10.95 & 0.18 & 1.40 & 1.64 \\
			&&$D$   & 16.62 & 0.12  & 2.14  &  2.50 \\
			\\
			\multirow{1}{*}{{J1738+0333}}
			&&$S$  &   10.71 & 0.20 & 1.51 & 1.62 \\
			\multirow{1}{*}{\bf 1.47  170.9}
			&&$A_1$& 10.66 & 0.21 & 1.49 & 1.60 \\
			&&$A_2$ & 10.59 & 0.21 & 1.48 & 1.59 \\
			&&$A_3$ & 11.42 & 0.19 & 1.74 & 1.86  \\
			&&$D$    & 16.08 & 0.14  &  2.30 & 2.47\\
			\\
			\multirow{1}{*}{{J0751+1807}}
			&&$S$   & 10.90 & 0.22 & 1.73 & 3.12 \\
			\multirow{1}{*}{\bf 1.64  287.4}
			&&$A_1$ & 10.86 & 0.22 & 1.72 & 3.11 \\
			&&$A_2$ & 10.82 & 0.23 & 1.72 & 3.10 \\
			&&$A_3$ & 11.87 & 0.21 & 2.10 & 3.80  \\
			&&$D$   & 15.45 & 0.16  & 2.39  & 4.32 \\
			\\
			\multirow{1}{*}{{J1012+5307}}
			&&$S$   & 10.90 & 0.25 & 1.88 & 2.25 \\
			\multirow{1}{*}{\bf 1.83  190.1}
			&&$A_1$ & 10.94 & 0.25 & 1.92 & 2.29 \\
			&&$A_2$ & 10.97 & 0.25 & 1.94 & 2.32 \\
			&&$A_3$ & 12.35 & 0.22 & 2.55 & 3.05  \\
			&&$D$   & 14.57 & 0.19  & 2.38  & 2.84 		
\end{tabular}}
	\label{table:pulsarmodels}
}
\end{table}

Finally, for slowly rotating objects with axial symmetry and a small angular velocity $\bar{\Omega}$ satisfying the condition $(\bar{\Omega} R)^2 \ll (M/R)$, we assume for the metric the ansatz \cite{hartle}
\be
\begin{split}
ds^2 = -e^{\nu(r)} dt^2 + \frac{1}{1-2 m(r)/r} dr^2 + r^2 d \Omega^2 \\
- 2 \omega(r) (r sin \theta)^2 dt d \phi
\end{split}
\en
for the interior problem, while outside the star the metric is given by the well-known Kerr solution \cite{kerr}, which for slowly rotating objects takes the simple form
\be
\begin{split}
ds^2=-\left( 1-\frac{2 M}{r} \right) dt^2 + \left( 1-\frac{2 M}{r} \right)^{-1} dr^2 + r^2 d \Omega^2 \\
- 2 \left( \frac{2 J}{r^3} \right) (r sin \theta)^2 dt d \phi
\end{split}
\en
where $J$ is the angular momentum of the rotating star~\cite{kokkotas1}.
 $J$ is given by 
\be
J=I\bar{\Omega}.
\en
 $I$ is the moment of inertia defined by 
\be
I = \frac{8 \pi}{3} \int_0^R dr r^4 (P+\epsilon) \left( \frac{e^\lambda}{e^\nu} \right)^{1/2} \left( \frac{\bar{\omega}}{\bar{\Omega}} \right)
\en
where we have defined the new function $\bar{\omega}=\bar{\Omega}-\omega$ satisfying the second order differential equation 
\be
e^{\nu-\lambda} \partial_r (e^{-\nu+\lambda} r^4 \partial_r \bar{\omega}) = 16 \pi r^4 (P+\epsilon) \bar{\omega}
\en
supplemented by the conditions
\begin{equation}
\bar{\omega} \rightarrow \Omega \; \; \; \; r \rightarrow \infty
\end{equation}
and
\begin{equation}
\frac{d \bar{\omega}}{dr}(0) = 0
\end{equation}
The second condition ensures regularity at the center, while the first condition ensures an asymptotically flat solution.

\begin{figure}[ht!]
	\centering
	\includegraphics[scale=0.7]{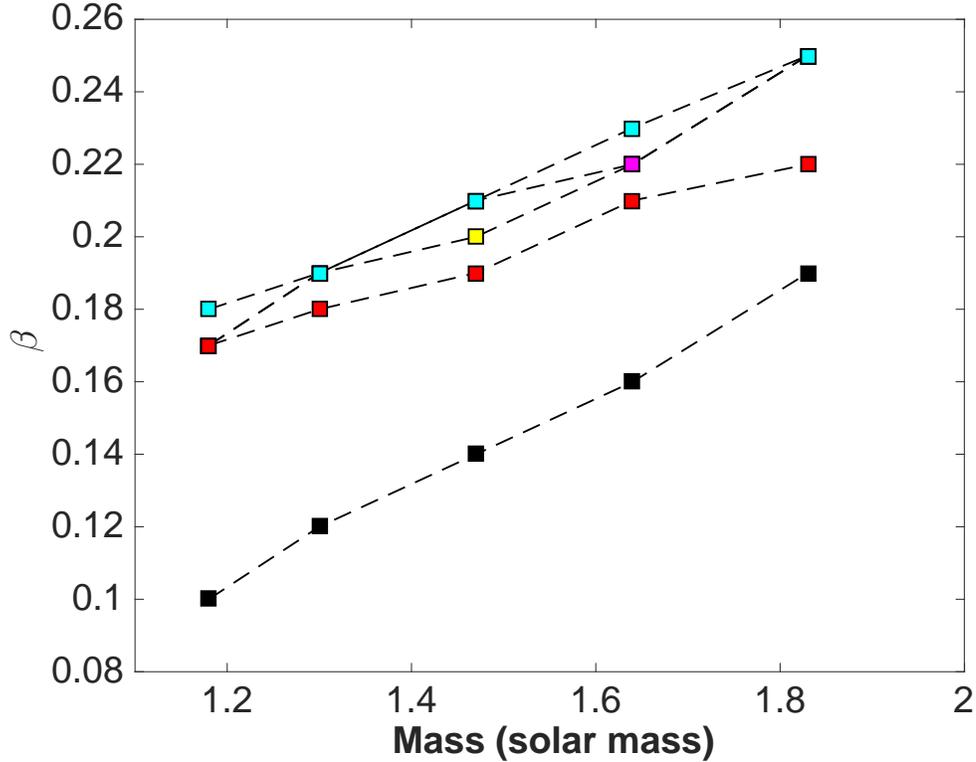}
	\caption{Compactness versus star mass for the five pulsars and the five models considered here. The colour scheme for the models is as follows: $S$ (0\% DM) in yellow, $A_1$ (5\% DM) in magenta, $A_2$ (11\% DM) in cyan, $A_3$ (25\%) in red and $D$ (100\% DM) in black. See Table~\ref{table:SetPulsars} for details.}
	\label{fig:1} 	
\end{figure}


\begin{figure}[ht!]
	\centering
	\includegraphics[scale=0.7]{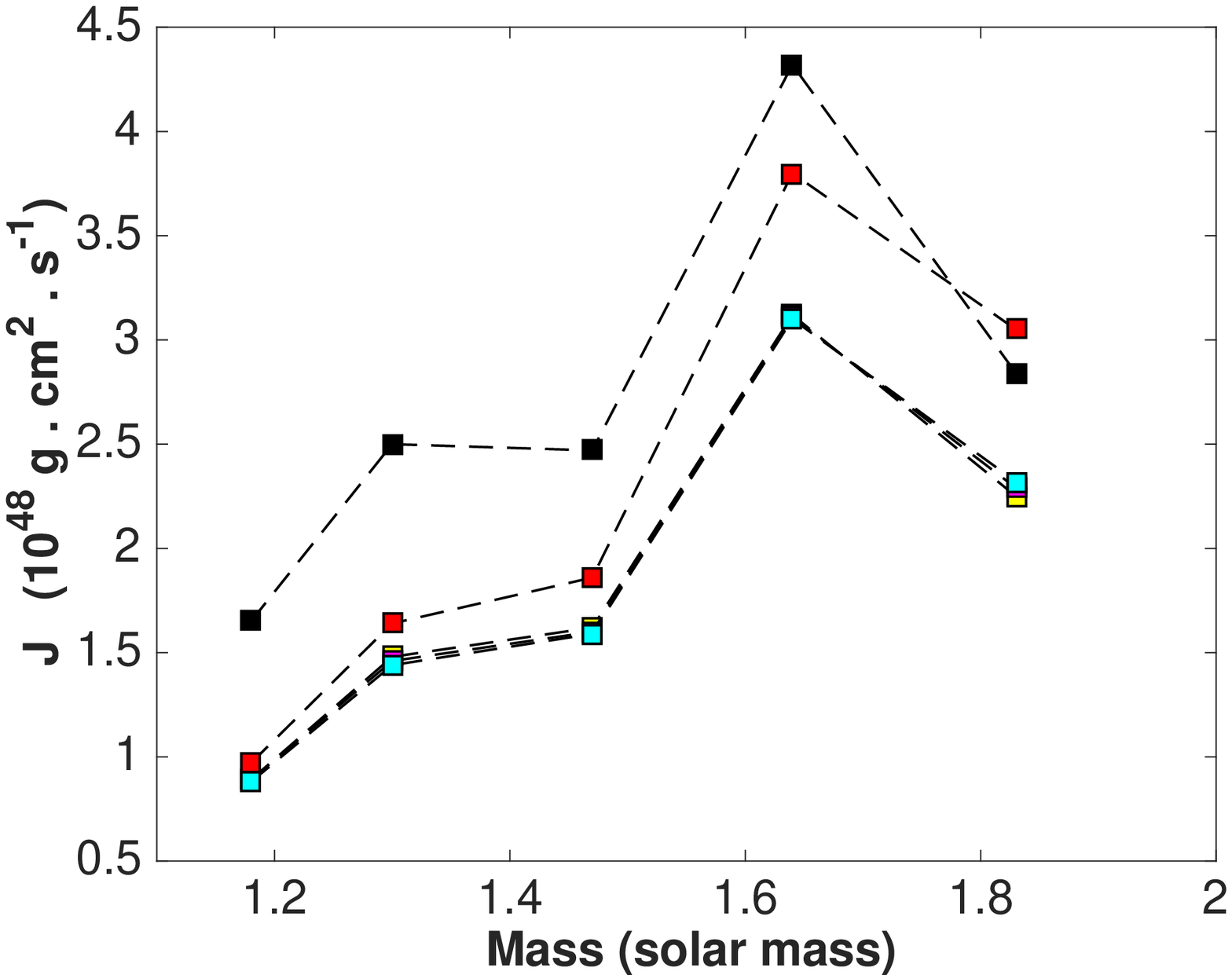}
	\caption{Angular momentum versus star mass for the five pulsars and the five models. The data points follow the same color scheme shown in Fig.~\ref{fig:1}.}
	\label{fig:2} 	
\end{figure}

\section{Properties of pulsars}

In the following we will compute pulsar properties such as the radius $R$, the compactness $\beta=M/R$, the moment of inertia $I$ and the angular momentum
$J$. The pulsars we have considered are shown in Table~\ref{table:SetPulsars}, while the five different models we have considered here are shown in Table ~\ref{table:SetDMstars}. As we have already mentioned, we have analyzed a purely strange star $S$, a purely condensed DM star $D$, and three admixed scenarios $A_1, A_2, A_3$ shown below:

\begin{equation}
A_1 \: \textrm{model} \rightarrow
\left\{
\begin{array}{lcl}
K = \frac{1.01}{B} \\
&
&
\\
 f = 0.09
\end{array}
\right.
\end{equation}


\begin{equation}
A_2 \: \textrm{model} \rightarrow
\left\{
\begin{array}{lcl}
K = \frac{0.46}{B} \\
&
&
\\
 f = 0.20
\end{array}
\right.
\end{equation}


\begin{equation}
A_3 \: \textrm{model} \rightarrow
\left\{
\begin{array}{lcl}
K = \frac{0.5}{B} \\
&
&
\\
 f = 0.9
\end{array}
\right.
\end{equation}


Within a given model/scenario, the appropriate value of the central pressure is required to reproduce the mass of the star, and after that all its properties, such as the radius or the DM mass fraction, can be unambiguously computed. In the model $A_1$ as we move from the lightest to the heaviest star the DM mass fraction varies from 4.3 \% to 5.6 \%, in $A_2$ varies from 9.6 \% to 12.2 \%, and in $A_3$ varies from 21.9 \% to 27.9 \%.

Our results are summarized in table~\ref{table:pulsarmodels}, and for better visualization we show the numerical results in Figures~\ref{fig:1}-~\ref{fig:2}.
There are two features that one immediately observes. First, regarding the radii of the stars, when there is a single fluid (model $S$ or $D$) things are clear, but when some DM is added (models $A_1, A_2, A_3$) the net result is the outcome of the competition between quarks and DM. Consequently, in some cases the radius decreases and in others increases. Generically we can say that the presence of DM tends to decrease the radius, unless there is a significant amount of DM, case in which the radius increases.

\medskip

The most important consequences of the presence of dark matter inside
a strange quark star are shown in Figures~\ref{fig:1} and ~\ref{fig:2}.
In a pure dark matter star (model D in tables 2 and 3) the compactness $\beta$ increases with the stellar mass. A larger amount of DM inside a strange star (model A3) behaves similar to a pure DM star. In this case $\beta$ decrease is a direct consequence of the increase of the radius of the star with the increase of the amount of dark matter inside the star (cf. Figure~\ref{fig:1}).
For instance, a strange quark star with total mass of $1.64\, M_\odot $ has a $\beta\approx 0.23$ (model $A_2$). If the same star has a larger amount of DM, its compactness factor is reduced.  

\smallskip
Furthermore, for a given star there is an increase of $I$ and $J$ as the DM mass fraction increases, although in the heaviest star with mass $M=1.83 M_{\odot}$ a change of this behaviour occurs in the last two cases. To understand this we need to rely on two factors (cf. Figure~\ref{fig:2}).: First, the moment of inertia is given by $I=a M R^2$, where $a$ is a numerical prefactor that is determined by the dynamics by solving the differential equation for $\omega(r)$. Second, for purely quark stars as well as for DM admixed quark stars the radius increases with the mass. On the contrary, due to the polytropic EoS in purely DM stars the radius decreases with the mass.
In the first four stars considered in this work, when we move from the model $A_3$ to the model $D$, although the prefactor $a$ decreases, overall the moment of inertia increases due to the large difference in the radii. However, in the heaviest star considered here, the difference in the radii is too small to compensate for the decrease in $a$, and therefore the moment of inertia of the model $D$ turns out to be smaller than that of the model $A_3$.

\smallskip
In summary, strange stars with or without dark matter have somehow quite similar properties. The compactness, momentum of inertia and the angular momentum vary with the increase of dark matter content in the star's interior. Nevertheless, if the dark matter content is very small, typically smaller than
10\%, its effects on the structure of the strange quark star is almost negligible, and 
the compactness, momentum of inertia and  the angular momentum do not vary much. However, if the amount of dark matter inside the strange quark star  is more than 25\% of its total mass, the previous quantities will vary significantly. In the case of the most massive pulsars the effect is very 
important. Moreover, in the case of  very massive strange stars, it will be difficult to infer the impact of DM on the structure of these stars, since a dark star has an $I$ and $J$ identical to the ones found for strange quark stars with 25\% of dark matter.

This theoretical work may suggest that among the population of pulsars in the Milky Way, there are many that could indeed be strange quark stars with dark matter.

\section{Conclusions}

In this work we have computed for the first time how the basic parameters of typical pulsars change for a new class of hypothetical compact objects. To be more specific, we have studied millisecond pulsars modelled as strange stars admixed with condensed DM. We have computed theoretically the stellar parameter values, such as radius, compactness, moment of inertia and angular momentum, of five observed pulsars with known masses and spin-periods. As far as the modelling of these observed pulsars is concerned, we have considered five different scenarios regarding the content of the compact star, namely a) purely strange star called $S$ in the text, b) purely condensed DM star called $D$, and c) three different scenarios of a strange star admixed with condensed DM (called $A_1$, $A_2$, $A_3$) depending on the numerical values of the $K$ constant in the EoS of DM as well as the $f$ parameter, see text. We have summarized our numerical results in tables and we have shown them graphically. The effects of DM are discussed.


\section*{Acknowlegements}

The authors thank the Funda\c c\~ao para a Ci\^encia e Tecnologia (FCT), Portugal, for the financial support to the Center for Astrophysics and Gravitation-CENTRA,  Instituto Superior T\'ecnico,  Universidade de Lisboa,  through the Grant No. UID/FIS/00099/2013.


\end{document}